\documentclass[aps,prb,preprint,showpacs,showkeys,floatfix]{revtex4}

\usepackage{graphicx,color}

\newcommand{\rev}[1]{\textcolor{black}{#1}}

\graphicspath{{figs/}}
\begin{document}

\title{Tunable negative Poisson's ratio in hydrogenated graphene}

\author{Jin-Wu Jiang}
    \altaffiliation{Corresponding author: jwjiang5918@hotmail.com}
    \affiliation{Shanghai Institute of Applied Mathematics and Mechanics, Shanghai Key Laboratory of Mechanics in Energy Engineering, Shanghai University, Shanghai 200072, People's Republic of China}

\author{Tienchong Chang}
    \affiliation{Shanghai Institute of Applied Mathematics and Mechanics, Shanghai Key Laboratory of Mechanics in Energy Engineering, Shanghai University, Shanghai 200072, People's Republic of China}

\author{Xingming Guo}
    \affiliation{Shanghai Institute of Applied Mathematics and Mechanics, Shanghai Key Laboratory of Mechanics in Energy Engineering, Shanghai University, Shanghai 200072, People's Republic of China}

\date{\today}
\begin{abstract}

We perform molecular dynamics simulations to investigate the effect of hydrogenation on the Poisson's ratio of graphene. It is found that the value for Poisson's ratio in graphene can be effectively tuned from positive to negative by varying the percentage of hydrogenation. Specifically, the Poisson's ratio decreases with the increase of the percentage of hydrogenation, and reaches a minimum value of -0.04 at the percentage of hydrogenation about 50\%. The Poisson's ratio starts to increase with a further increase of the percentage of hydrogenation. The appearance of minimum negative Poisson's ratio in the hydrogenated graphene is attributed to the suppression of the hydrogenation-induced ripples during stretching of graphene. \rev{Our results demonstrate that hydrogenation is a valuable approach for the tuning of the Poisson's ratio from positive to negative in graphene.}

\end{abstract}
\maketitle
\pagebreak

\section{Introduction}

\rev{Materials with negative Poisson's ratio (NPR) are allowed by the classical elasticity theory, which sets a range of $-1< \nu < 0.5$ for the Poisson's ratio in an isotropic three-dimensional material.\cite{LandauLD} These NPR Materials are termed as auxetic materials by Evans.\cite{EvansKE1991Endeavour} Auxetic materials possess some novel properties such as enhanced toughness and enhanced sound or vibration absorption,\cite{LipsettAW1988jasm} so the auxeticity has attracted intense research interest since Lakes observed the auxetic phenomenon in the permanently compressed foam structure in 1987.\cite{LakesRS1987sci} Since then, great efforts have been devoted to finding or designing new auxetic materials.\cite{RothenburgL1991nat,LakesR1993adm,BaughmanRH1993nat,EvansKE2000adm,YangW2004jmsci,RaviralaN2007jms,LethbridgeZAD2010am,BertoldiK2010am,GreavesGN2011nm,AldersonK2012pssb,clausenAM2015} 
It turns out that the auxeticity is actually quite common in man-made structures or natural materials. For example, Baughman et al. found that 69\% of the cubic elemental metals have intrinsic NPR along a non-axial direction.\cite{BaughmanRH1998nat} This non-axis NPR phenomenon was previously obtained by Milstein and Huang for some face centered cubic structures.\cite{MilsteinF1979prb}
}

\rev{While most previous works are about the auxeticity of bulk materials,\cite{LakesR1993adm,EvansKE2000adm,YangW2004jmsci,GreavesGN2011nm} there have been few examples demonstrating the NPR phenomenon for low-dimensional nanomaterials in recent years. The NPR was found to be intrinsic for some nanomaterials, including the single-layer black phosphorus\cite{JiangJW2014bpnpr} or the few-layer orthorhombic arsenic.\cite{HanJ2015ape} As a result of the large surface to volume ratio in nanoplates, the surface-induced phase transformation can cause obvious NPR for several metal nanoplates.\cite{HoDT2014nc} Yao et al. investigated possible NPR in carbon nanotubes by virtual modifications of structural parameters or bonding strength using the valence force constant model.\cite{YaoYT2008pssb}
}

As one of the most widely studied low-dimensional nanomaterials, graphene has also been found to exhibit NPR phenomenon. Specifically, the Poisson's ratio for graphene can be driven into the negative regime by thermally induced ripples at high temperatures above 1700~K.\cite{ZakharchenkoKV} Large numbers of vacancy defects\cite{GrimaJN2015adm,hoPSSB2016a} for bulk graphene or compressive edge stress-induced warping in graphene ribbons are two different mechanisms to achieve the NPR in graphene.\cite{JiangJW2016npr_fbc} A more recent work reveals the NPR to be an intrinsic material property for single-layer graphene with strain above 6\%.\cite{JiangJW2016npr_intrinsic}

Graphene's peculiar electronic properties are basically related to the $\pi$-electron. It is possible to alter electronic properties of graphene by attaching atoms or moleculars to graphene, which can remove $\pi$-electrons. The reversible hydrogenation (graphene linked to hydrogen) is one of such technique,\cite{SofoJO2007prb,EliasDC2009sci} which leads to $sp^2$-$sp^3$ hybridization of carbon atoms attached by hydrogen atoms. The reversibility of the hydrogenation process enables graphene to be a promising hydrogen storage material.\cite{BoukhvalovDW2008prb} It has also been shown that the hydrogenation is useful for tuning various properties in graphene, such as the electronic band gap,\cite{BalogR2009nm,HabererD2010nl} magnetic properties,\cite{ZhouJ2009apl,ZhouJ2009nl} thermal conductivity,\cite{PeiQX2011carbon} mechanical properties,\cite{PeiQX2010carbon,LeenaertsO2010prb,Cadelano2012prb} and etc.

\rev{There are some first-principles calculations on the Poisson's ratio in graphane (the fully hydrogenated graphene),\cite{CadelanoE2010prb,TopsakalM2010apl,PengQ2013pccp,AnsariR2015ssc} and a weak NPR phenomenon was obtained in one of the fully hydrogenated graphene structures.\cite{CadelanoE2010prb,ColomboL2011ppp} However, these first-principles calculations are based on small unit cells, so the effects from the hydrogenation-induced ripples\cite{ZhuS2014acsn} can not be included. Furthermore, these first-principles calculations are for fully hydrogenated graphene, while the dependence of the Poisson's ratio on the percentage of hydrogenation for partially hydrogenated graphene is still unclear, which will be the main focus of the present work.}

In this paper, we investigate the Poisson's ratio for the hydrogenated graphene with percentage of hydrogenation $p\in[0, 1]$. The Poisson's ratio decreases linearly with increasing the percentage of hydrogenation in a lower percentage range $p\in[0,0.2]$, and reaches a minimum value at $p\approx 0.5$. The Poisson's ratio starts to increase with further increasing percentage of hydrogenation. In particular, the Poisson's ratio is tuned into negative by hydrogenation with percentage $p\in[0.2, 0.8]$. The effects of hydrogenation on the Poisson's ratio are interpreted based on the suppression of the hydrogenation-induced ripples during the stretching of graphene.

\section{Structure and simulation details}
\begin{figure}[tb]
  \begin{center}
    \scalebox{1}[1]{\includegraphics[width=8cm]{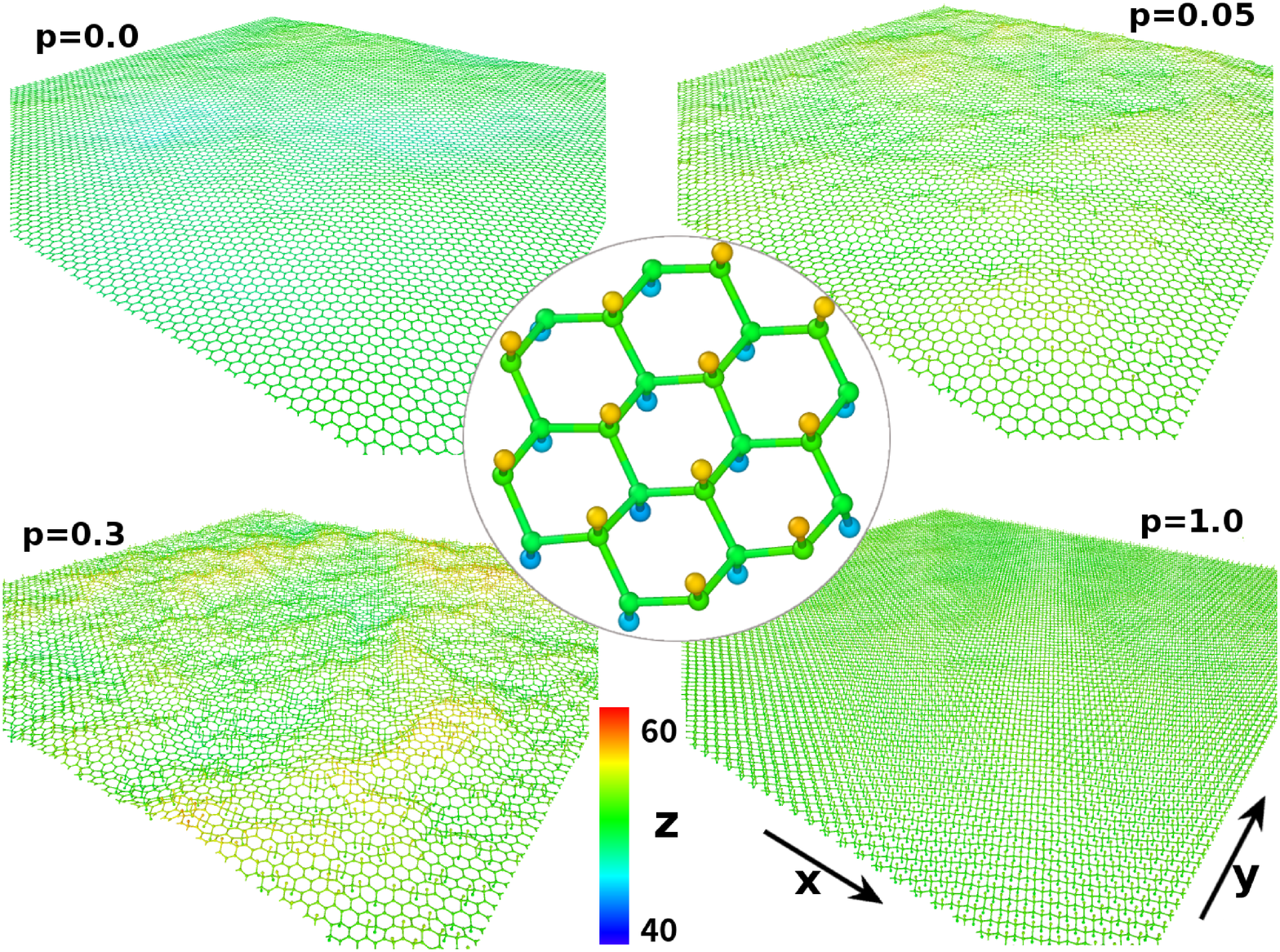}}
  \end{center}
  \caption{(Color online) Structure for hydrogenated graphene of size $200\times 200$~{\AA} at room temperature. The left top panel shows the pure graphene (i.e., percentage of hydrogenation $p=0.0$). The other three panels show the hydrogenated graphene with $p=0.05$, 0.3, and 1.0. The central inset shows the chairlike hydrogenation pattern for the fully hydrogenated graphene with $p=1.0$, where hydrogen atoms are bonded to carbon atoms on both sides of the plane in an alternating manner. The colorbar is with respective to the z-coordinate of each atom.}
  \label{fig_cfg_hydrogen}
\end{figure}

The configuration of hydrogenated graphene is shown in Fig.~\ref{fig_cfg_hydrogen}, where hydrogen atoms are attached to graphene of size $200\times 200$~{\AA} at room temperature. The left top panel is for pure graphene (i.e., percentage of hydrogenation $p=0.0$). The other three panels show the hydrogenated graphene with the percentage of hydrogenation $p=0.05$, 0.3, and 1.0. The formula of the hydrogenated graphene is CH for the fully hydrogenated graphene with $p=1.0$, i.e., all carbon atoms are attached to one hydrogen atom. The hydrogenation is in the stable chairlike form as displayed in the central inset, where hydrogen atoms are bonded to carbon on both sides of graphene in an alternating manner.\cite{SofoJO2007prb} \rev{We focus on this chairlike hydrogenation for graphene, as it is the most stable form with the lowest ground-state energy, although the other two hydrogenated structures are also thermodynamically accessible according to first-principles calculations.\cite{CadelanoE2010prb} For the partially hydrogenated graphene with hydrogenation percentage of $p$, we actually construct it in a reverse manner; i.e., the partially hydrogenated structure is obtained by randomly removing $(1-p)$ percentage of hydrogen atoms from the chairlike fully hydrogenated graphene. In this way, we can maintain the same chairlike hydrogenation style for all partially hydrogenated graphene, so that the computations and discussions for the hydrogenation effect can be self-consistent. In this way, the isotropic property of the chairlike hydrogenation can be well kept for any partially hydrogenated graphene.}

\begin{figure}[tb]
  \begin{center}
    \scalebox{1}[1]{\includegraphics[width=8cm]{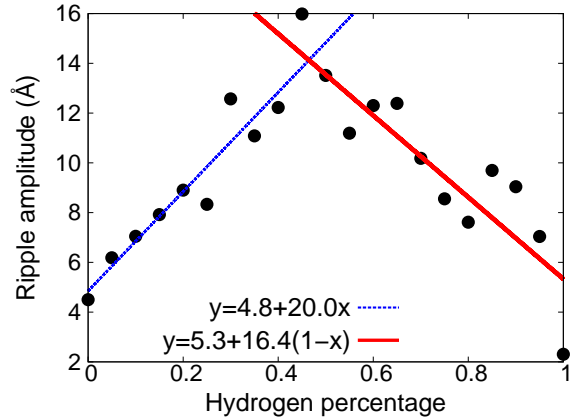}}
  \end{center}
  \caption{(Color online) The amplitude for the hydrogenation-induced ripple versus the percentage of hydrogenation at room temperature.}
  \label{fig_amplitude_hydrogen}
\end{figure}

Hydrogenation induces some ripples in graphene, which has been discussed in previous works.\cite{ZhuS2014acsn} There are more ripples with the increase of the percentage of hydrogenation $p=0.0$, 0.05, and 0.3. However, the number of ripple decreases with further increasing percentage of hydrogenation for $p>0.5$. There is almost no ripple in the fully hydrogenated graphene with percentage of hydrogenation $p=1.0$, because the fully hydrogenated graphene regains the periodic crystal structure. We note that there are actually some thermal-induced ripples in the pure graphene or the fully hydrogenated graphene. However, the amplitudes of the thermal-induced ripple are almost neglectable as compared with the hydrogenation-induced ripples. \rev{The thermal-induced ripples correspond to the vibration of the long wave phonon modes in graphene. Consequently, these ripples are more smooth and have longer wave length, as shown by the left top panel in Fig.~\ref{fig_cfg_hydrogen}. The hydrogenation-induced ripples are caused by the local curvature resulting from the formation of the C-H chemical bonds between hydrogen and graphene. As a result, these ripples have smaller wave length, as shown by the right top and left bottom panels in Fig.~\ref{fig_cfg_hydrogen}.} Fig.~\ref{fig_amplitude_hydrogen} shows the amplitudes of ripples versus the percentage of hydrogenation. The hydrogenation-induced ripples have the largest amplitude in the hydrogenated graphene with a moderate percentage of hydrogenation at $p\approx0.5$.

The carbon-carbon interactions are described by the \rev{second generation Brenner potential},\cite{brennerJPCM2002} which has been widely used to study the mechanical response of graphene.\cite{MoY2009nat} The structure is stretched in the x (armchair)-direction while graphene is allowed to be fully relaxed in the y (zigzag)-direction, using molecular dynamics simulations. Periodic boundary conditions are applied in both x and y-directions. Both edges in the x-direction are clamped during the stretching of graphene in the x-direction; i.e., carbon atoms in the edge regions at $\pm$x ends are forbidden to move in the z-direction. The standard Newton equations of motion are integrated in time using the velocity Verlet algorithm with a time step of 1~{fs}. Simulations are performed using the publicly available simulation code LAMMPS~\cite{PlimptonSJ}, while the OVITO package is used for visualization~\cite{ovito}.

\section{Results and discussions}

\begin{figure}[tb]
  \begin{center}
    \scalebox{1}[1]{\includegraphics[width=8cm]{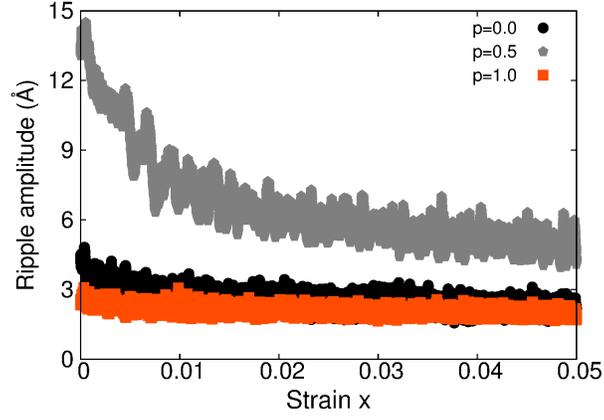}}
  \end{center}
  \caption{(Color online) The strain dependence for the amplitude of the ripples in the hydrogenated graphene during the stretching of graphene at room temperature.}
  \label{fig_amplitude_T300}
\end{figure}

The structure of the partially hydrogenated graphene in Fig.~\ref{fig_cfg_hydrogen} looks similar as a crumpled sheet of paper with many ripples. Inspired by the fact that ripples in the crumpled paper will be suppressed by stretching, we first investigate the evolution of the amplitude for the ripple during the stretching of graphene in Fig.~\ref{fig_amplitude_T300}. Indeed, similar as the crumpled paper, the amplitude of the hydrogenation-induced ripple decreases with the increase of the applied tensile strain, indicating a suppression of the ripples during the stretching process. Such suppression of ripples may result in the expansion of the hydrogenated graphene in the lateral direction by stretching it along the longitudinal direction, which is similar as the expansion of a crumpled sheet of paper during its tensile deformation process. That is the suppression of the ripples serves as an underlying mechanism for possible NPR in the hydrogenated graphene.

\begin{figure}[tb]
  \begin{center}
    \scalebox{1}[1]{\includegraphics[width=8cm]{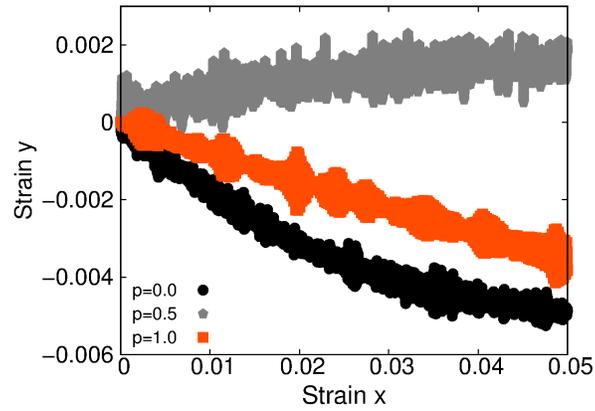}}
  \end{center}
  \caption{(Color online) The resultant strain $\epsilon_y$ versus the applied strain $\epsilon_x$ at room temperature for the hydrogenated graphene with percentage of hydrogenation $p=0.0$, 0.5, and 1.0.}
  \label{fig_strain_T300}
\end{figure}

To calculate the Poisson's ratio, we show in Fig.~\ref{fig_strain_T300} the resultant strain in the y-direction for a given applied strain in the x-direction in hydrogenated graphene at room temperature. The Poisson's ratio is obtained as the slope of the curve in a small strain range $\epsilon_{x}\in [0, 0.01]$. The Poisson's ratio is sensitive to the percentage of hydrogenation. For instance, the Poisson's ratio has the value of 0.16, -0.04, and 0.07 corresponding to the percentage of hydrogenation 0.0, 0.5, and 1.0, respectively. \rev{We note that the hydrogenation can cause distinct effect on the strain-strain curves shown in Fig.~\ref{fig_strain_T300} even for a small strain (eg. 0.01) applied in the x-direction. The applied strain is thus limited to be smaller than 0.05 in the present work.}

\begin{figure}[tb]
  \begin{center}
    \scalebox{1}[1]{\includegraphics[width=8cm]{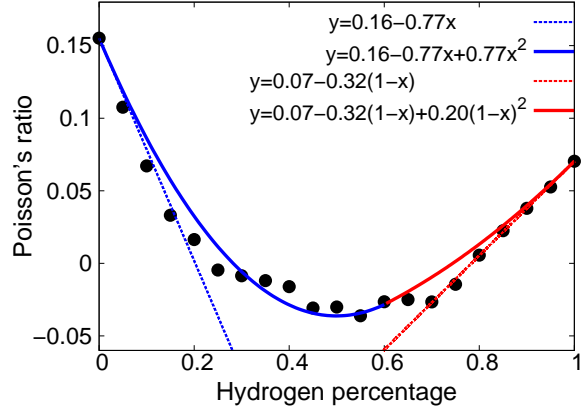}}
  \end{center}
  \caption{(Color online) The Poisson's ratio versus the percentage of hydrogenation for graphene of size $200\times 200$~{\AA} at room temperature.}
  \label{fig_poisson_hydrogen}
\end{figure}

Fig.~\ref{fig_poisson_hydrogen} shows the dependence of the Poisson's ratio on the percentage of hydrogenation in the hydrogenated graphene at room temperature. There is a distinct valley point in the curve; i.e., a minimum negative Poisson's ratio of $\nu=-0.04$ at the percentage of hydrogenation $p\approx 0.5$. \rev{The negative value of -0.04 is not very large as compared with the NPR obtained in bulk materials, which can approach to the theoretical lower limit of -1.\cite{LakesRS1987sci} However, this value may be large enough to cause visible effects for nanomaterials like graphene, considering the ultra-high atomic resolution of the measurement on the nanoscale level. We note that similar NPR value of -0.07 was reported due to thermal ripples at temperatures above 1700~K.\cite{ZakharchenkoKV}.} The Poisson's ratio decreases with the increase of the percentage of hydrogenation for $p<0.5$, but the Poisson's ratio starts to increase with further increasing percentage of hydrogenation for $p>0.5$. There are obvious linear relations between the Poisson's ratio and the percentage of hydrogenation for $p\in[0,0.2]$ or $p\in[0.8, 1.0]$. A linear fitting in these two percentage ranges gives $\nu=0.16-0.77p$ for $p\in[0, 0.2]$, and $\nu=0.07-0.32(1-p)$ for $p\in[0.8, 1.0]$. The dependence of the Poisson's ratio on the percentage of hydrogenation deviates from linear for moderate percentage range $p\in[0.2, 0.8]$. A nonlinear fitting leads to $\nu=0.16-0.77p+0.77p^2$ for $p\in[0.2, 0.5]$, and $\nu=0.07-0.32(1-p)+0.2(1-p)^2$ for $p\in[0.5, 0.8]$, where the coefficients for the linear terms are kept the same as the linear fitting for $p\in[0, 0.2]$ or $p\in[0.8, 1.0]$.

We will now illustrate that the minimum Poisson's ratio in the hydrogenated graphene with $p\approx 0.5$ is closely related to the hydrogenation-induced ripples. For pure graphene, the Poisson's ratio is a positive value of $\nu_0=0.16$. For hydrogenated graphene, the hydrogenation-induced ripple will be suppressed by stretching graphene as shown in Fig.~\ref{fig_amplitude_T300}, which results in the expansion of the hydrogenated graphene during stretching (like a crumpled sheet of paper). As a result, the suppression of hydrogenation-induced ripples leads to the reduction of Poisson's ratio. This is the reason for the decrease of Poisson's ratio for $p\in [0, 0.5]$. The hydrogenated graphene with $p\in [0.5, 1.0]$ can be regarded as a disordered system originating from the fully hydrogenated graphene. The percentage of disorder is $(1-p)$. Similarly, the Poisson's ratio of the disordered hydrogenated graphene will decrease with increasing the percentage of disorder $(1-p)$. That is the Poisson's ratio increases with increasing percentage of hydrogenation for $p\in [0.5, 1.0]$.

\begin{figure}[tb]
  \begin{center}
    \scalebox{1}[1]{\includegraphics[width=8cm]{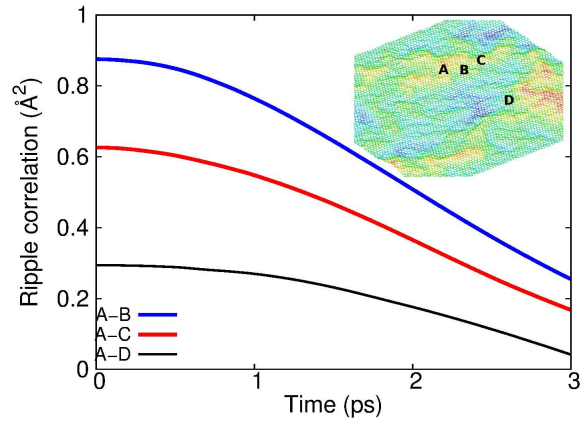}}
  \end{center}
  \caption{(Color online) The correlation between the amplitudes of the ripple A and the ripples B, C, and D. The correlation at time $t$ is obtained as the ensemble average $C_{ij}(t)=<A_{i}(0)A_{j}(t)>$, with $A_{i}$ and $A_{j}$ as the amplitudes for ripples i and j. The correlation $C_{ij}$ decreases with increasing distance between ripples i and j.}
  \label{fig_correlation}
\end{figure}

\rev{We regard the hydrogenated graphene as a hybrid structure of the fully hydrogenated domain of percentage $p$ and the pure graphene domain of percentage $(1-p)$. As a result, the Poisson's ratio value contributed by the fully hydrogenated domain is $p\nu_r$, while the Poisson's ratio value contributed by the pure graphene domain is $(1-p)\nu_0$ with $\nu_0=0.16$. The actual Poisson's ratio for the partially hydrogenated graphene can be obtained as $\nu=(1-p)\nu_0+p\nu_r=\nu_0+(\nu_r-\nu_0)p$. This is indeed a linear function, which explains the above numerical and fitting results. For larger percentage of hydrogenation $p\in[0.2, 0.5]$, there are more ripples, so the correlations between two neighboring ripples become important. Fig.~\ref{fig_correlation} shows that the correlation becomes stronger for two closer ripples. The correlations between neighboring ripples lead to the nonlinear feature in the hydrogenation percentage dependence of the Poisson's ratio.}

For the hydrogenated graphene with $p\in[0.5, 1.0]$, we can consider the system as a derivative of the fully hydrogenated graphene crystal with the percentage of disorder $(1-p)$; i.e., hydrogen atoms are removed by a percentage of $(1-p)$ from the fully hydrogenated graphene. Similarly, we find that the Poisson's ratio in $p\in[0.8, 1.0]$ can be fitted to a linear function $\nu=0.07-0.32(1-p)$, which gives $\nu_r=-0.25$ as the Poisson's ratio contributed by the ripples. For $p\in[0.5, 0.8]$, the dependence of the Poisson's ratio on the percentage of hydrogenation is a nonlinear function.

\begin{figure}[htpb]
  \begin{center}
    \scalebox{1}[1]{\includegraphics[width=8cm]{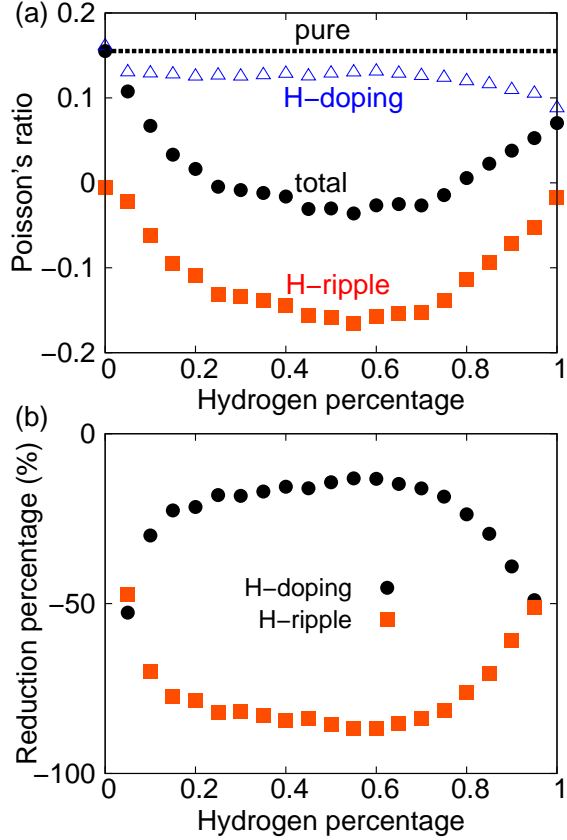}}
  \end{center}
  \caption{(Color online) Two separate effects (doping and ripple) from the hydrogenation on the Poisson's ratio in hydrogenated graphene. (a) The total Poisson's ratio (black dots) for the hydrogenated graphene is the summation of the Poisson's ratio with doping effect (blue open triangles) and the Poisson's ratio corresponding to the ripple effect (red filled squares). The Poisson's ratio for pure graphene, $\nu_0=0.16$, is also displayed by the dashed horizontal line. (b) The reduction percentage (with reference to $\nu_0$) of the Poisson's ratio by the doping effect and the ripple effect.}
  \label{fig_poisson_decompose}
\end{figure}

In the above, we have demonstrated the overall effects from the hydrogenation to the Poisson's ratio for graphene. Actually, the hydrogenation causes two individual effects on graphene. First, the C-H chemical bonds formed by hydrogenation can influence the original C-C chemical bonds in graphene. This is a particular type of doping effect. This doping effect can influence mechanical properties for graphene, including the Poisson's ratio. Second, hydrogenation induces some obvious ripples in graphene. The suppression of the ripples during the tensile deformation of the hydrogenated graphene makes negative contribution to the Poisson's ratio value. We can differentiate separate contributions from these two effects. To do so, we perform artificial simulations with the z-coordinates for all atoms fixed, which effectively freezes the ripples during stretching. The ripple effect is excluded in these artificial simulations, while the doping effect on the Poisson's ratio is left. We thus get the Poisson's ratio that is solely affected by the doping effect (denoted by $\nu_{\rm doping}$). The Poisson's ratio corresponding to the ripple effect can be obtained by $\nu_{\rm tot}-\nu_{\rm doping}$, with $\nu_{\rm tot}$ as the actual (total) Poisson's ratio for the hydrogenated graphene with both effects considered. These results are comparatively shown in Fig.~\ref{fig_poisson_decompose}~(a). Obviously, the valley point for the Poisson's ratio originates from the ripple effect. We also compare the percentage of the reduction in the Poisson's ratio (with reference to $\nu_{0}$) contributed by the doping or ripple effects in Fig.~\ref{fig_poisson_decompose}~(b), which clearly shows major contribution from the ripple effect.

\rev{Actually, doping is a chemical process, so it can affect the chemistry of the C-C bonds in graphene. As an evidence, the length of the C-C bond is 1.42~{\AA} in pure graphene, and the bond length is increased to 1.52~{\AA} by doping with -H in the fully hydrogenated graphene.\cite{SofoJO2007prb} The increase of bond length reflects that the C-C bond is weakened.\cite{HarrisonWA1999elementary} That is, the interaction between carbon atoms will be slightly weakened by hydrogen doping. In other words, the force constant $K_b$ for bond stretching is decreased, while the force constant for angle bending $K_{\theta}$ is not affected. The Poisson's ratio for graphene is related to these two force constants as,\cite{ChangT2003jmps}
\begin{eqnarray}
\nu = 1 - \frac{24}{K_b a^2/K_{\theta} + 18}.
\label{eq_nu_vffm}
\end{eqnarray}
According to Eq.~(\ref{eq_nu_vffm}), the Poisson's ratio will be reduced by the decrease of the force constant $K_b$. This is the underlying mechanism for the doping induced reduction of the Poisson's ratio found in the above. We also found that the ripple effect is to cause obvious reduction for the Poisson's ratio of graphene. There is an entropy mechanism underlying this effect. More specifically, the flattening of the geometrical ripples due to stretching can reduce the phase space. The structure tries to expand in the unstretched direction for the purpose of minimizing entropy energy.\cite{ZakharchenkoKV} This abnormal deformation leads to obvious reduction of the Poisson's ratio in graphene.
}

\rev{Before concluding, we note that we have demonstrated the hydrogenation-induced ripple effect on the Poisson's ratio in graphene. These ripples are caused by the local curvature induced by the chemical bonding C-H. Other atoms/groups like -O or -OH can also form chemical bonds with graphene. It is thus expected that similar reduction in the Poisson's ratio should occur, although the magnitude of the reduction may show some quantitative differences due to different chemical bonds.}

\section{Conclusion}
In conclusion, we have investigated the effect of hydrogenation on the Poisson's ratio for graphene through molecular dynamics simulations. We found a minimum negative Poisson's ratio for the hydrogenated graphene. Specifically, the Poisson's ratio decreases with increasing percentage of hydrogenation in the percentage range [0, 0.5], and the Poisson's ratio starts to increase with further increase of the percentage of hydrogenation. The phenomenon of a minimum negative Poisson's ratio is caused by the suppression of the hydrogenation-induced ripples during the stretching of the hydrogenated graphene.

\section*{Acknowledgements}
JWJ thanks Ning Wei for discussions on the usage of LAMMPS. The work is supported by the Recruitment Program of Global Youth Experts of China, the National Natural Science Foundation of China (NSFC) under Grant Nos. 11504225, 11472163, 11425209, and the start-up funding from Shanghai University.

%

\end{document}